\documentclass{article}

\usepackage{latexsym,amssymb,amsfonts,amsmath,amsthm}
\usepackage{comment}
\usepackage{tikz}

\newtheorem{theorem}{Theorem}

\newtheorem{corollary}{Corollary}

\newtheorem{remark}{Remark}

\numberwithin{theorem}{section}
\numberwithin{lemma}{section}
\numberwithin{corollary}{section}
\numberwithin{proposition}{section}
\numberwithin{remark}{section}

\newcommand{\bs}[1]{\boldsymbol{#1}}

\newcommand{\dist}{{\rm dist}}

\usepackage{color}

\title{A convergence time of Grover walk \\ on regular graph to stationary state}
\author{
Ayaka Ishikawa$^1$, 
Sho Kubota$^1$, 
Etsuo Segawa$^2$\footnote{Corresponding author: segawa-etsuo-tb@ynu.ac.jp}\\
{\small $^1$ Department of Applied Mathematics, Faculty of Engineering, Yokohama National University} \\
{\small $^2$ Graduate School of Environment and Information Sciences, Yokohama National University} \\
{\small$^{1,2}$ Hodogaya, Yokomhama, 240-8501, Japan}
}
\date{}

\begin{document}

\maketitle

\begin{small}
\par\noindent
{\bf Abstract}. 
We consider a quantum walk model on a finite graph which has an interaction with the outside. Here a quantum walker from the outside penetrates the graph and also a  quantum walker in the graph goes out to the outside at every time step. This dynamics of the quantum walk converges to a stationary state. In this paper, we estimate the speed of the convergence to the stationary state on the $\kappa$-regular graph with the uniformly inserting of the inflow to the graph. We show that larger degree of the regular graph makes the convergence speed of this  quantum walk model slower.  
\end{small}

\section{Introduction}
Interesting phenomena and applications such as the cut off phenomenon and electric circuit  derive from fruitful mathematical results on the mixing time and hitting time of random walks~\cite{DS,Diaconis,LevinPeres}. 
Such phenomena are based on the well-known fact that random walks on finite graphs converge to stationary states.
Quantum walks can be explained as a quantum analogue of random walks~\cite{Meyer,AmbainisEtAl}. Quantum walks sometimes accomplish so called quantum speed up in the quantum search on graphs, which is quadratic speed up comparing with a classical search algorithm~\cite{Ambainis2003,Childs, Portugal}.

However, every absolute value of the eigenvalue of the time evolution operator of the quantum walk is unit because of the unitarity of the time evolution. Then the mixing time of the quantum walk is infinite in general if we apply directly this unitary operator as the time evolution. Thus there needs to be some schemes to construct the unitary dynamical system which converges to a stationary state~\cite{Brassard}. For example, instead of the stationary state directly, the Ces{\'a}ro summation of the finding probability has been considered, and also the convergence to a stationary state is accomplished in the quantum search by the time dependent unitary time evolution~\cite{Grover,YLC}. Another interesting approach is that the domain of the time evolution is expanded from the square summable space to the uniformly bounded space~\cite{FelHil1,FelHil2}. 
In this scheme, the time evolution operator can be independent of the time step, which means that this time evolution can be simply explained as a natural dynamics. 
In this paper, we adopt such a model introduced by \cite{FelHil1,FelHil2} because the time evolution of the corresponding random walk, which has a finite mixing time, is also time independent in general. Now let us explain briefly our quantum walk model treated here. First we set a semi-infinite graph from the original finite graph by connecting the finite number of semi-infinite paths to some vertices of the original graph. Secondly, we set the local dynamics of the quantum walk at each vertex so that it moves free on the
tails. Finally, we set the initial state, which is an uniformly bounded but no longer square summable, so that quantum walkers penetrate into the internal graph from the tails at every time step. 
Since the dynamics on the tails is free, once a quantum walker in the internal graph goes out to the tails, then it never goes back to the internal graph. Such a quantum walker can be regarded as the outflow. Then if we restrict our attention to this dynamics to the internal graph, then the walk restricted to the internal graph is reduced to a finite open system which interacts with the outside. It is shown that such a dynamical system converges to a stationary state as a fixed point~\cite{FelHil1,FelHil2,HS}. 

There are studies on the characterization of such stationary state~\cite{HSS2,HSS3}. For example, the stationary state of the Grover walk with the constant frequency of the inflow can be described by the electric circuit and the energy in the internal graph, comforatability, can be expressed by some graph geometry. As a sequential studies on this quantum walk model, we are also interested in the aspect of the behavior closing to the stationary state. Then, in this paper, we focus on the speed of the convergence to the stationary state which corresponds to the mixing time of random walks. As a first trial, we insert the constant inflow to every vertex of $\kappa$-regular graph and estimate when the system converges to the stationary state; that is, the convergence speed of quantum walk. We obtain the expression of the convergence speed of quantum walks using the number of vertices and the degree of the graph. This expression implies that the higher degree of the graph delays the speed of the convergence. Then our result leads to the following quite counter-intuitive statement from the viewpoint of random walks:  in all the regular graphs with the vertex number $N$, the graph which has the fastest convergence speed of the quantum walk is the 
cycle graph, while the graph which has the slowest convergence speed is the complete graph.

This paper is organized as follows. 
In section~2, the setting of the graph and also the quantum walk are devoted. 
In section~3, we present the main theorem for the convergence speed of this quantum walk. 
In section~4, we give the proof of the main theorem. 
Finally, we give the summary and discussion in section~5. 
\section{Settings}
\subsection{Graph with tails}
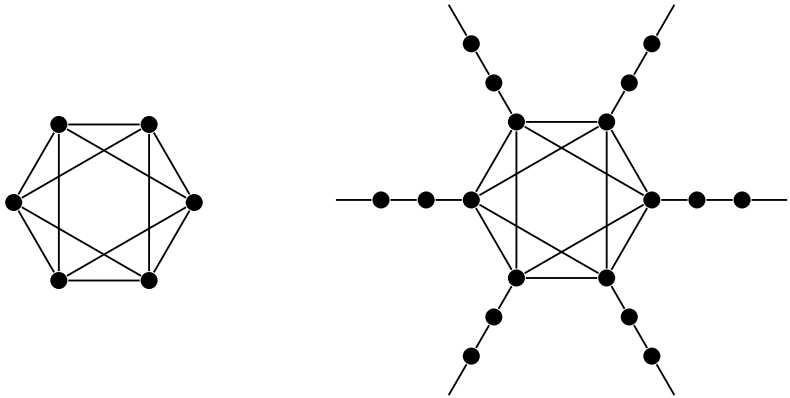
\begin{figure}[ht]
\begin{center}
\begin{tikzpicture}
[scale = 0.6, line width = 0.7pt,
v/.style = {circle, fill = black, inner sep = 0.8mm}]
\node[v] (1) at (2, 0) {};
\node[v] (2) at (1, 1.73) {};
\node[v] (3) at (-1, 1.73) {};
\node[v] (4) at (-2, 0) {};
\node[v] (5) at (-1, -1.73) {};
\node[v] (6) at (1, -1.73) {};
\draw (1) -- (2) -- (3) -- (4) -- (5) -- (6) -- (1);
\draw (1) -- (3) -- (5) -- (1);
\draw (2) -- (4) -- (6) -- (2);
\end{tikzpicture}
$\qquad \qquad$
\raisebox{-14.25mm}{
\begin{tikzpicture}
[scale = 0.6, line width = 0.7pt,
v/.style = {circle, fill = black, inner sep = 0.8mm}]
\node[v] (1) at (2, 0) {};
\node[v] (2) at (1, 1.73) {};
\node[v] (3) at (-1, 1.73) {};
\node[v] (4) at (-2, 0) {};
\node[v] (5) at (-1, -1.73) {};
\node[v] (6) at (1, -1.73) {};
\draw (1) -- (2) -- (3) -- (4) -- (5) -- (6) -- (1);
\draw (1) -- (3) -- (5) -- (1);
\draw (2) -- (4) -- (6) -- (2);
\node[v] (11) at (3, 0) {};
\node[v] (111) at (4, 0) {};
\draw (1) -- (11) -- (111) -- +(1,0);
\node[v] (22) at (1.5, 2.595) {};
\node[v] (222) at (2, 3.463) {};
\draw (2) -- (22) -- (222) -- +(0.5, 0.865);
\node[v] (33) at (-1.5, 2.595) {};
\node[v] (333) at (-2, 3.463) {};
\draw (3) -- (33) -- (333) -- +(-0.5, 0.865);
\node[v] (44) at (-3, 0) {};
\node[v] (444) at (-4, 0) {};
\draw (4) -- (44) -- (444) -- +(-1, 0);
\node[v] (55) at (-1.5, -2.595) {};
\node[v] (555) at (-2, -3.463) {};
\draw (5) -- (55) -- (555) -- +(-0.5, -0.865);
\node[v] (66) at (1.5, -2.595) {};
\node[v] (666) at (2, -3.463) {};
\draw (6) -- (66) -- (666) -- +(0.5, -0.865);
\end{tikzpicture}
}
\caption{Graph with tails} \label{K01}
\end{center}
\end{figure}

Let $G=(V,A)$ be a connected and finite symmetric digraph. 
Here a symmetric digraph is the digraph in which every arc $a\in A$ has the inverse arc $\bar{a}\in A$.  
Then letting $t(a)$ and $o(a)$ be the terminal and origin vertices of $a\in A$, we have $t(a)=o(\bar{a})$ and $o(a)=t(\bar{a})$. 
We prepare for the same number of semi-infinite paths as the vertices
and join each root of the semi-infinite length path to every vertex of $G$. Such a resulting infinite graph like a ``hedgehog" is denoted by $\tilde{G}=(\tilde{V},\tilde{A})$. 
The semi-infinite length tails are called the tails. 
We take a graph as an example on the left hand side of Figure~\ref{K01}.
On the right hand side, the graph with tails is illustrated.
The degree of $u\in V$ is denoted by $d(u)$; that is, $d(u)=|\{a\in A \;|\; t(a)=u\}|$, while the degree of $u\in \tilde{V}$ is denoted by $\tilde{d}(u)$.
Then for any $u\in V$, we have  $\tilde{d}(u)=d(u)+1$. 

\subsection{Time evolution of Grover walk}
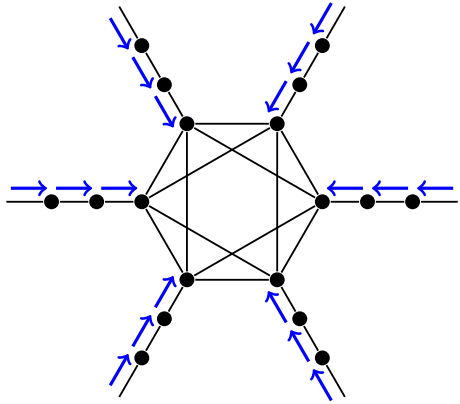
\begin{figure}[ht]
\begin{center}
\begin{tikzpicture}
[scale = 0.6, line width = 0.7pt,
v/.style = {circle, fill = black, inner sep = 0.7mm}]
\node[v] (1) at (2, 0) {};
\node[v] (2) at (1, 1.73) {};
\node[v] (3) at (-1, 1.73) {};
\node[v] (4) at (-2, 0) {};
\node[v] (5) at (-1, -1.73) {};
\node[v] (6) at (1, -1.73) {};
\draw (1) -- (2) -- (3) -- (4) -- (5) -- (6) -- (1);
\draw (1) -- (3) -- (5) -- (1);
\draw (2) -- (4) -- (6) -- (2);
\node[v] (11) at (3, 0) {};
\node[v] (111) at (4, 0) {};
\draw (1) -- (11) -- (111) -- +(1,0);
\draw[blue, line width = 1.2pt, ->] (2.9,0.3) to (2.1,0.3);
\draw[blue, line width = 1.2pt, ->] (3.9,0.3) to (3.1,0.3);
\draw[blue, line width = 1.2pt, ->] (4.9,0.3) to (4.1,0.3);
\node[v] (22) at (1.5, 2.595) {};
\node[v] (222) at (2, 3.463) {};
\draw (2) -- (22) -- (222) -- +(0.5, 0.865);
\draw[blue, line width = 1.2pt, ->] (1.2,2.676) to (0.8,1.984);
\draw[blue, line width = 1.2pt, ->] (1.7,3.541) to (1.3,2.849);
\draw[blue, line width = 1.2pt, ->] (2.2,4.406) to (1.8,3.714);
\node[v] (33) at (-1.5, 2.595) {};
\node[v] (333) at (-2, 3.463) {};
\draw (3) -- (33) -- (333) -- +(-0.5, 0.865);
\draw[blue, line width = 1.2pt, ->] (-2.7, 4.071) to (-2.3, 3.379);
\draw[blue, line width = 1.2pt, ->] (-2.2, 3.206) to (-1.8, 2.514);
\draw[blue, line width = 1.2pt, ->] (-1.7, 2.341) to (-1.3, 1.649);
\node[v] (44) at (-3, 0) {};
\node[v] (444) at (-4, 0) {};
\draw (4) -- (44) -- (444) -- +(-1, 0);
\draw[blue, line width = 1.2pt, ->] (-2.9,0.3) to (-2.1,0.3);
\draw[blue, line width = 1.2pt, ->] (-3.9,0.3) to (-3.1,0.3);
\draw[blue, line width = 1.2pt, ->] (-4.9,0.3) to (-4.1,0.3);
\node[v] (55) at (-1.5, -2.595) {};
\node[v] (555) at (-2, -3.463) {};
\draw (5) -- (55) -- (555) -- +(-0.5, -0.865);
\draw[blue, line width = 1.2pt, ->] (-1.7, -2.341) to (-1.3, -1.649);
\draw[blue, line width = 1.2pt, ->] (-2.2, -3.206) to (-1.8, -2.514);
\draw[blue, line width = 1.2pt, ->] (-2.7,-4.071) to (-2.3, -3.379);
\node[v] (66) at (1.5, -2.595) {};
\node[v] (666) at (2, -3.463) {};
\draw (6) -- (66) -- (666) -- +(0.5, -0.865);
\draw[blue, line width = 1.2pt, ->] (1.2, -2.676) to (0.8,-1.984);
\draw[blue, line width = 1.2pt, ->] (1.7,-3.541) to (1.3,-2.849);
\draw[blue, line width = 1.2pt, ->] (2.2,-4.406) to (1.8,-3.714);
\end{tikzpicture}
\caption{The initial state $\Psi_0$} \label{K02}
\end{center}
\end{figure}

For a countable set $\Omega$, the vector space, whose standard basis are labeled by $\Omega$, is denoted by $\mathbb{C}^{\Omega}$. The standard basis are expressed by the delta functions $\{\delta_\omega^{\Omega}\}_{\omega\in \Omega}$. 
The total vector state of the quantum walk is denoted by $\mathbb{C}^{\tilde{A}}$. The time evolution operator on $\mathbb{C}^{\tilde{A}}$ is described as follows: 
\[ (U\Psi)(a)=\frac{2}{\tilde{d}(o(a))}\sum_{\substack{b \in \tilde{A} \\ t(b) = o(a)}} \Psi(b)-\Psi(\bar{a}) \]
for any $a\in \tilde{A}$ and $\Psi\in \mathbb{C}^{\tilde{A}}$. 
Let $\Psi_t$ be the $t$-th interation of quantum walk; that is, $\Psi_{t+1}=U\Psi_t$. 
Set $\{a_1,\dots,a_{\tilde{d}(u)}\}:=\{a\in \tilde{A} \;|\;t(a)=u \}$. Then the local time evolution is described by  
\[ \begin{bmatrix} \Psi_{t+1}(\bar{a}_1) \\ \vdots \\ \Psi_{t+1}(\bar{a}_{\tilde{d}(u)}) \end{bmatrix} = \mathrm{Gr}(\tilde{d}(u))\begin{bmatrix} \Psi_{t}(a_1) \\ \vdots \\ \Psi_{t}(a_{\tilde{d}(u)}) \end{bmatrix}.  \]
Here $\mathrm{Gr}(d)$ is the $d$-dimensional Grover matrix such that 
\[ \mathrm{Gr}(d)=\frac{2}{d}J_d-I_d, \]
where $J_d$ is the all-ones matrix. 
This means that the Grover matrix presents the local scattering on each vertex at each time step; the transmitting and reflecting amplitudes are given by $2/\tilde{d}(u)$ and $2/\tilde{d}(u)-1$, respectively.
Note that if the degree of every vertex of the tail is $2$, the diagonal term of the Grover matrix becomes $0$. Then the quantum walk on the tails behaves free. The initial state is set so that quantum walkers penetrate into the internal graph at every time step as follows. 
\begin{equation}\label{eq:initial} \Psi_0(a)=
\begin{cases}
1 & \text{: $a\in$ arc set of tails, $\dist (t(a);G)<\dist (o(a);G)$} \\
0 & \text{: otherwise.}
\end{cases} 
\end{equation}
The support of the initial state is all the arcs towards the graph.
Taking the graph in Figure~\ref{K01} as an example,
the initial state $\Psi_0$ is illustrated in Figure~\ref{K02}.
The entries of $\Psi_0$ corresponding to the blue arcs are $1$,
and the others are $0$.

\subsection{Total variance of quantum walker}
Let $\chi:\mathbb{C}^{\tilde{A}}\to\mathbb{C}^{A}$ be the restriction to the internal; that is, 
\[ (\chi \Psi)(a) = \Psi(a) \;\text{for any $a\in A$ and $\Psi\in\mathbb{C}^{\tilde{A}}$}\]
and its adjoint $\mathbb{C}^{A}\to\mathbb{C}^{\tilde{A}}$ 
\[ (\chi^* \psi)(a) =\begin{cases} \psi(a) & \text{: $a\in A$} \\0 & \text{: $a\notin A$} \end{cases} \]
for any $a\in \tilde{A}$ and $\psi\in \mathbb{C}^{A}$. 
Then a matrix expression for $\chi$ is 
\[ \chi\cong [\; I_{A}\;|\;\bs{0}\; ] \]
which is the $|A|\times \infty $ matrix under the decomposition of $\tilde{A}=A\cup (\tilde{A}\setminus A)$. Then it is easily to see that $\chi\chi^*$ is the identity operator of $\mathbb{C}^{A}$ while $\chi^*\chi$ is the projection operator onto $\mathbb{C}^{A}$. 
Note that the inflow penetrating into the internal graph $G$ at time $t$ ($t\geq 1$) is expressed by 
$\rho_t:=\chi U(I_{\tilde{A}}-\chi^*\chi)\Psi_{t-1}$.
It is easy to verify that
\[\rho(a) = \frac{2}{k+1} \]
for any $a \in A$.
By the definition of the initial state $\Psi_0$, the inflow $\rho_t$ is independent of the time step; that is, $\rho:=\rho_1=\rho_t$ for any $t\geq 1$. 
Since the dynamics on the tails are trivial, we will concentrate on the internal graph as follows. Let $\psi_t:=\chi \Psi_t$, and $E_{PON}:=\chi U \chi^*$ which is the truncated matrix with respect to the internal graph. Then we have~\cite{HS} 
\begin{equation}\label{eq:TE}
\psi_{t+1}=E_{PON}\psi_t+\rho,\;\;\psi_0=0. 
\end{equation}
It is know that $\psi_t$ converges to a stationary state $\psi_\infty$ as $t\to\infty$~\cite{HS}. 
The relative probability measure on $V$ is defined by 
\[ \mu_t(u)=\sum_{\substack{a \in A \\ t(a)=u}} |\psi_t(a)|^2. \]
Note that $\sum_{u\in V}\mu_t(u)\neq 1$ in general.  
The stationary measure of $\mu_t$ is denoted by $\mu_\infty$; that is, 
\[ \mu_\infty(u)=\sum_{\substack{a \in A \\ t(a)=u}}|\psi_\infty(a)|^2. \]
The total variance of quantum walk between $\mu,\nu\in \mathbb{C}^{V}$ with $\mu(u),\nu(u)\geq 0$ for any $u\in V$ is defined by 
\[ || \mu-\nu ||_{QTV}:=\max_{V'\subset V} |\mu(V')-\nu(V')| \]
in this paper. 
The total variance of quantum walk can be reexpressed by 
\begin{equation}\label{eq:TVE} ||\mu-\nu||_{QTV}=\frac{1}{2}\left(\sum_{u\in V} |\mu(u)-\nu(u)|+|\mathcal{E}_{\mu}-\mathcal{E}_{\nu}|\right), 
\end{equation}
where $\mathcal{E}_{h}$ ($h\in \{\mu,\nu\}$) is the comfortability~\cite{HSS3}; that is, 
\[ \mathcal{E}_h=\sum_{u\in V} h(u). \]
We give the proof in Appendix.  
We are interested in the speed of the convergence to the stationary state through this variance. 
\section{Main theorem}
Let the total variance of quantum walk at time $t$ with the inflow (\ref{eq:initial}) be $d_t$ such that 
\[d_t:=|| \mu_t-\mu_\infty ||_{QTV}. \]
We set the convergence speed of quantum walk by 
\[ t_*(\theta):=\min\{t\in\mathbb{N}\;:\;d_t<e^{-\theta}\} \]
for $\theta\geq 0$. 
Then we obtain the following theorem. 
\begin{theorem}\label{thm:main}
Let $G=(V,A)$ be a connected and $\kappa$-regular graph with $N$ vertices and $\theta\geq 0$. 
The convergence speed of quantum walk with the inflow (\ref{eq:initial}) is estimated as follows. 
\[ \frac{1}{\log\frac{\kappa+1}{\kappa-1}}\left( \log \kappa N+\theta  \right)\leq t_*(\theta)\leq \frac{1}{\log\frac{\kappa+1}{\kappa-1}} \log 2\kappa N . \]
\end{theorem}
Note that if the larger the degree of the graph is, the slower the convergence speed of quantum walk is,  which is a counter intuitive behavior from the view point of the random walk. 
Let us see an example of Theorem~\ref{thm:main} in the follwing. 
Let $\Gamma^k(N)$ ($1\leq k\leq \lfloor N/2 \rfloor$) be the circulant graph with the vertex set $\{0,1,\dots,N-1\}$, where vertices $i$ and $j$ are adjacent if and only if 
\[i-j\in\{\pm 1,\pm 2,\dots,\pm k\}.\]
Note that $\Gamma^1(N)$ is the cycle graph while $\Gamma^{N-1}(N)$ is the complete graph with the vertex number $N$. 
If $\Gamma^\kappa(N)$ is the cycle graph ($k=1$), then the convergence speed is $t_*(\theta)\in O(\log N)$, which is quite faster than that of the simple lazy random walk while if $\Gamma^\kappa(N)$ is the complete graph ($k=N-1$), then $t_*(\theta)\in O(N\log N)$, which is slower than that of the simple random walk:
for simple random walk case, the convergence speed of the lazy random walk on the cycle graphs is $O(N^2)$ and that on the complete graph is $O(1)$, respectively. 
\begin{table}[htbp]
\begin{center}
\begin{tabular}{r|cc}
& Cycle graph & Complete graph \\\hline
Random walk  & $O(N^2)$~\cite{Diaconis,LevinPeres} & $O(1)$ \\
Quantum walk & $O(\log N)$ & $O(N \log N)$
\end{tabular}
\end{center}
\caption{Comparison of the mixing time of random walk and the convergence speed of quantum walk}
\end{table}
In general, it is easy to see the case of quantum walk in the following corollary. 
\begin{corollary}
Let the degree of the circulant graph $\Gamma^\kappa(N)$ be $k \in O(N^{\alpha})$ with $0\leq \alpha \leq 1$. Then the convergence speed of quantum walk on $\Gamma_k(N)$ with the initial state (\ref{eq:initial}) is estimated by
\[ t_*(\theta)\in O(N^\alpha \log N). \]
\end{corollary}
If we set a uniform initial state in the random walk, then we cannot see any dynamical time evolution. Thus the initial state of the mixing time for the random walk is set to have the slowest convergence time. On the other hand, we can obtain a kind of the duality relation between thee mixing time and the convergence time of the quantum walk by setting a natural initial state of the quantum walk, which is the uniform.   
\section{Proof of Theorem~\ref{thm:main}}
\begin{proof}
Assume that the original graph $G=(V,A)$ is a $\kappa$-regular graph which is symmetric and connected.  
Let us set important three operators $M$, $K$, $S$ in the following. 
The adjacency matrix of $G$ is denoted by  $M: \mathbb{C}^V\to\mathbb{C}^V$; that is,
\[ (Mf)(u)=\sum_{\substack{a \in A \\ t(a)=u}}f(o(a)) \]
for any $f\in\mathbb{C}^V$ and $u\in V$. The matrix expression is  
\[ (M)_{u,v}=\begin{cases} 1 & \text{: $u$ and $v$ are adjacent in $G$,}\\
0 & \text{: otherwise.}
\end{cases} \]
The operator $K:\mathbb{C}^{A}\to \mathbb{C}^V$ is a weighted incidence matrix between each arc and its terminal vertex defined by
\[ (K\psi)(a)=\frac{1}{\sqrt{\kappa+1}}\sum_{\substack{a \in A \\ t(a)=u}}\psi(a)  \]
for any $\psi\in \mathbb{C}^A$. 
The matrix representation of $K=(K_{u,a})_{u\in V,\;a\in A}$, which is the $|V|\times |A|$-matrix, is given by \[ (K)_{u,a}=\begin{cases} 1/\sqrt{\kappa+1} & \text{: $t(a)=u$,} \\ 0 & \text{: otherwise.} \end{cases} \]
Let $S$ be the shift operator such that $(S\psi)(a)=\psi(\bar{a})$ for any $\psi\in \mathbb{C}^A$; that is, 
\[(S)_{a,b}=\begin{cases}1 & \text{: $t(b)=o(a)$,}\\ 0 & \text{: otherwise.}\end{cases}\]
It is easy to check that 
\begin{align}
    E_{PON} &= S (2K^*K-I_{A}); \label{eq:Epon}\\
    KK^{*} &= \frac{\kappa}{\kappa+1} I_V;\label{eq:kk} \\
    KSK^{*} &= \frac{1}{\kappa+1} M.\label{eq:adj}
\end{align}
Note that the inflow $\rho\in\mathbb{C}^{A}$ in (\ref{eq:TE}) can be expressed by 
\begin{equation}\label{eq:rho} \rho=\frac{2}{\sqrt{\kappa+1}}SK^*\bs{1}_{V}. \end{equation}
Let $L$ be the $|A|\times 2|V|$ matrix such that $L:=[\;K^*\;|\;SK^* \;]$.
Let $E_{GON}$ be the $2|V|\times 2|V|$-matrix defined by
\[E_{GON}=\begin{bmatrix} 0 & -I_{V} \\ \frac{\kappa-1}{\kappa+1}I_V & \frac{2}{\kappa+1}M \end{bmatrix}.  \]
Then using (\ref{eq:Epon})--(\ref{eq:adj}), we have
\begin{equation}\label{eq:PONGON}
    E_{PON}L=LE_{GON}.
\end{equation}
This equation means that the time evolution of the quantum walk restricted to the internal graph, $E_{PON}$, can be switched to that of the $E_{GON}$, which is based on the adjacency matrix. We are relatively more familiar with the treatment of the adjacency matrix than that of quantum walk operator directly. Then using (\ref{eq:PONGON}), we will try to express the  $t$-th iteration of the quantum walk, $\psi_t$, in the literature of the adjacency matrix in the following.  Assume the eigenvalues of the adjacency matrix $M$ are $\kappa=\lambda_1>\lambda_2\geq \cdots \geq \lambda_{N}\geq -\kappa$ and the corresponding eigenprojections are denoted by $P_1, P_2\cdots, P_N$, respectively. 
Then $E_{GON}$ can be decomposed into
\begin{equation}\label{eq:decom}
E_{GON}=\sum_{j=1}^{|V|}\left(\begin{bmatrix} 0 & -1 \\ \frac{\kappa-1}{\kappa+1} & \frac{2\lambda_j}{\kappa+1}\end{bmatrix}\otimes P_j\right). 
\end{equation}
We put $\bs{q}:=(2/\sqrt{\kappa+1})\bs{1}_V$ and 
\[ \Lambda_j:=\begin{bmatrix} 0 & -1 \\ \frac{\kappa-1}{\kappa+1} & \frac{2\lambda_j}{\kappa+1}\end{bmatrix}. \]
Note that 
\begin{equation}\label{eq:q}
    P_j\bs{q}=\delta_{1,j}\bs{q}
\end{equation}
because $\bs{q}\in \ker(\kappa-M)$ and $\dim \ker(\kappa-M)=1$ by the Perron-Frobenius theorem. 
Then it holds that 
\begin{align}
    \psi_t &=(I_{A}+E_{PON}+\cdots+E_{PON}^{t-1})\;\rho  &\text{ ( By (\ref{eq:TE}) )} \notag\\
    &=(I_{A}+E_{PON}+\cdots+E_{PON}^{t-1}) L\begin{bmatrix} \bs{0} \\  \bs{q}\end{bmatrix} &\text{( By (\ref{eq:rho}) )}\notag \\
    &= L(I_{2|V|}+E_{GON}+\cdots+E_{GON}^{t-1}) \begin{bmatrix} \bs{0} \\  \bs{q}\end{bmatrix} &\text{( By (\ref{eq:PONGON}) )}\notag \\
    &= L\;\sum_{j}(I_2+\Lambda_j+\cdots+\Lambda_j^{t-1})\begin{bmatrix}0\\1\end{bmatrix}\otimes P_j\bs{q} &\text{( By (\ref{eq:decom}) )}\label{eq:general} \\
    &= L\;\left\{(I_2+\Lambda_1+\cdots+\Lambda_1^{t-1})\begin{bmatrix}0\\1\end{bmatrix}\otimes \bs{q}\right\} &\text{( By (\ref{eq:q}) )}. \label{eq:medium}
\end{align}
The eigenvalues of $\Lambda_1$ are 
$1$, $(\kappa-1)/(\kappa+1)$ and corresponding eigenprojections are 
\[ Q_1=\frac{1}{2}\begin{bmatrix} -(\kappa-1) & -(\kappa+1) \\\kappa-1 & \kappa+1 \end{bmatrix},\; Q_2=\frac{1}{2}\begin{bmatrix} \kappa+1 & \kappa+1 \\ -(\kappa-1) & -(\kappa-1) \end{bmatrix}, \]
respectively. Then 
the matrix $\Lambda_1$ can be decomposed into 
\[ \Lambda_1=Q_1+\frac{\kappa-1}{\kappa+1}Q_2. \]
 Note that $K^*\bs{q}=SK^*\bs{q}$ because $\bs{q}$ is a constant function on $\mathbb{C}^V$. 
Since $L=[1\;0]\otimes K^*+[0\;1]\otimes SK^*$, equation (\ref{eq:medium}) can be rewritten by 
\begin{align}
    \psi_t 
    &= t\{\;(Q_1)_{1,2}+(Q_1)_{2,2}\;\} K^*\bs{q} \notag\\
    &\qquad\qquad +\frac{1-\left(\frac{\kappa-1}{\kappa+1}\right)^t}{1-\frac{\kappa-1}{\kappa+1}} \{\;(Q_2)_{1,2}+(Q_2)_{2,2}\;\} K^*\bs{q} \label{eq:psit1} \\
    &= \frac{1-\left(\frac{\kappa-1}{\kappa+1}\right)^t}{1-\frac{\kappa-1}{\kappa+1}} \{\;(Q_2)_{1,2}+(Q_2)_{2,2}\;\} K^*\bs{q} \notag \\
    &= \left\{ 1-\left(\frac{\kappa-1}{\kappa+1}\right)^t \right\}\bs{1}_{A}. \label{eq:psit}
\end{align} 
\noindent From (\ref{eq:psit}), we immediately obtain 
\begin{equation}\label{eq:tv}
    d_t=||\;\mu_\infty-\mu_t\;||_{QTV}=\left(\frac{\kappa-1}{\kappa+1}\right)^t\left\{ 2-\left( \frac{\kappa-1}{\kappa+1} \right)^t \right\}\kappa |V|. 
\end{equation}
The RHS of (\ref{eq:tv}) can be bounded by 
\[ \left(\frac{\kappa-1}{\kappa+1}\right)^t\kappa |V|<d_t<\left(\frac{\kappa-1}{\kappa+1}\right)^t2\kappa |V|. \]
Taking the logarithm, we obtain the desired conclusion. 
\end{proof}
\begin{remark}
Since $\log[\kappa N]/\log[(\kappa+1)/(\kappa-1)]$ is monotone increasing with respect to $\kappa$, then in all the simple regular graph, which has no multiple edges and self-loops, with the vertex number $N$, the graph which is the fastest convergence speed of the quantum walk is the $2$-regular graph, i.e., the cycle graph, while the graph which is the slowest convergence speed is the $N (N-1)/2$-regular graph, i.e., the complete graph. 
\end{remark}
\begin{remark}
The reason that the coefficient of the blow up term in (\ref{eq:psit1}); $(Q_1)_{1,2}+(Q_1)_{2,2}$, is vanished may be explained in more general argument by \cite{HS}. 
\end{remark}
As a by-product of the proof of this theorem, we obtain the following theorem which derives from (\ref{eq:psit}) directly.
\begin{theorem}\label{thm:tthiteretion}
For any connected and $\kappa$-regular graph $G=(V,A)$, the $t$-th iteration of quantum walk with the initial state (\ref{eq:initial}) is described by \begin{equation}
    \psi_t= \left\{1-\left( \frac{\kappa-1}{\kappa+1}\right)^t\right\}\bs{1}_{A_0}.
\end{equation}
\end{theorem}
\noindent This theorem tells us that the amplitude at every time step can be described by a constant value with respect to each arc at every time step. This derives from the argebraic relation (\ref{eq:PONGON}) between the truncated time evolution $E_{PON}$ and the extended adjacency matrix $E_{GON}$,  and also the common property of the $\kappa$-regular connected graphs that the adjacency matrix has the simple eigenvalue $\kappa$ and its eigenvector is a constant function.   \\
\section{Summary and discussion}
We estimated a convergence speed to the stationary state of the Grover walk on $\kappa$-regular graph with the in- and out-flows. If the inflow is uniform, then the convergence speed is estimated by using the degree $\kappa$ and the number of vertices $N$ (see Theorem~\ref{thm:main}).  
This result derives from the algebraic relation between quantum walk and random walk (\ref{eq:PONGON}) and the common property of the adjacency matrix that the eigenvector of the maximal eigenvalue is a constant vector.
We believe that the expression for $\psi_t$ in Theorem~\ref{thm:tthiteretion} is not trivial. Such a constant inflow denoted by $\bs{q}$ gives a simple analysis on the estimation of the convergence speed, however it also gives somewhat a monotonous description of $\psi_t$ unfortunately. Then if we improve the notion of the convergence speed of quantum walk by changing the worst inflow $\bs{q}'$ for the speed from the uniform inflow $\bs{q}$, such a convergence speed seems to be more reasonable and be expected to tell us some interesting geometric information. We also expect that this study becomes a stepping stone to work on the cut off phenomena of quantum walks. We leave these in this paper as our interesting future's works. 

\noindent \\
\noindent{\bf Acknowledgement:}
We would like to thank to Norio Konno, Sarato Takahashi, Yusuke Higuchi and Takuya Ohwa for useful comments. 
A.I. acknowledges financial supports from the Grant-in-Aid for JSPS Research Fellow (Grant No. JP20J20590).
S.K. acknowledges financial supports from the Grant-in-Aid for JSPS Research Fellow (Grant No. 20J01175). E.S. acknowledges financial supports from the Grant-in-Aid of
Scientific Research (C) Japan Society for the Promotion of Science (Grant No.~19K03616) and Research Origin for Dressed Photon.
\appendix
\section{Expression for total variance of quantum walk}
We prove (\ref{eq:tv}) based on \cite{LevinPeres}. 
Let us set $B:=\{u\in V\;|\; \mu(u)>\nu(u)\}$. Then for any subset $V'\subset V$, it holds that  
\begin{align*}
    \mu(V')-\nu(V') &=\mu(V'\cap B)-\nu(V'\cap B)+\mu(V'\cap B^{c})-\nu(V'\cap B^{c})\\
    &\leq \mu(A\cap B)-\nu(A\cap B) \\
    &\leq \mu(B)-\nu(B),
\end{align*}
where $B^c=V\setminus B$. 
In the same way, we have 
\[ \nu(V')-\mu(V')\leq \nu(B^{c})-\mu(B^{c}). \]
Noting that $\mu(B^c)=\mathcal{E}_\mu-\mu(B)$ and $\nu(B^c)=\mathcal{E}_\nu-\nu(B)$, we obtain
\[ \nu(V')-\mu(V')\leq \mu(B)-\nu(B)-\Delta_{\mu,\nu}, \]
where $\Delta_{\mu,\nu}:=\mathcal{E}_{\mu}-\mathcal{E}_\nu$. If $\Delta_{\mu,\nu}>0$, then 
\begin{align*}
    |\mu(V')-\nu(V')| &\leq \mu(B)-\nu(B) \\
    &= \frac{1}{2}\left\{\; (\mu(B)-\nu(B))+ (\nu(B^c)-\mu(B^c)+\Delta_{\mu,\nu}) \;\right\} \\
    &= \frac{1}{2}\left\{ \sum_{u\in V} |\mu(u)-\nu(u)|+\Delta_{\mu,\nu} \right\}. 
\end{align*}
On the other hand, if $\Delta_{\mu,\nu}<0$, then
\begin{align*}
    |\mu(V')-\nu(V')| &\leq \mu(B)-\nu(B)-\Delta_{\mu,\nu} \\
    &= \frac{1}{2}\left\{\; (\mu(B)-\nu(B))+ (\nu(B^c)-\mu(B^c)+\Delta_{\mu,\nu}) \;\right\}-\Delta_{\mu,\nu} \\
    &= \frac{1}{2}\left\{ \sum_{u\in V} |\mu(u)-\nu(u)|-\Delta_{\mu,\nu} \right\}. 
\end{align*}
Then we have 
\[ |\mu(V')-\nu(V')|\leq \frac{1}{2}\left\{ \sum_{u\in V} |\mu(u)-\nu(u)|+|\Delta_{\mu,\nu}|\right\}. \]
Taking the maximum with respect to $V'$ in the LHS, we obtain 
\[ \max_{V'\subset V}|\mu(V')-\nu(V')|=\frac{1}{2}\left\{ \sum_{u\in V} |\mu(u)-\nu(u)|+|\Delta_{\mu,\nu}|\right\}, \]
which is the desired conclusion. 

\end{document}